\documentclass[letter,traditabstract]{aa}
\usepackage{graphicx}
 \usepackage{lscape}
 \usepackage{colortbl}
 \usepackage{natbib}
 \usepackage{verbatim} 
  \usepackage{comment} 
\usepackage{ulem}
\usepackage{url}
 \bibpunct{(}{)}{;}{a}{}{,}
\usepackage{txfonts}
\begin{document}
\input psfig.tex
\def\simgt{\stackrel{>}{{}_\sim}}
\def\simlt{\stackrel{<}{{}_\sim}}

\titlerunning{Age spread at $z\sim2.2$}

   \title{The extreme synchronicity of stellar ages of red  
   galaxies in the JKCS\,041 cluster at $z=2.2$} 
   \author{S. Andreon} 

   \institute{
             Observatorio Astronomico di Brera, via Brera 28, 20121, Milan, Italy\\
             \email{stefano.andreon@brera.inaf.it} 
}

   \date{Received --, 2011; accepted --, 2011}

\abstract{Above redshift $z\sim 1.4$, we known little or nothing
about the stellar ages of red galaxies
in clusters, yet at these high redshifts important changes 
are predicted by
current renditions of galaxy formation models embedded in 
the standard hierarchical paradigm of structure formation.
Red-sequence galaxies in the cluster JKCS\,041 at $z=2.2$
show a tight distribution in colour that
indicates a star formation history that is highly synchronized 
across galaxies. Specifically, we measure a 
spread in stellar age of $160\pm 30$ Myr, in marked disagreement with
the current understanding of how massive red galaxies form in clusters, 
i.e. if they are produced somewhat stochastically in
merging episodes that sometimes involve gas, hence star formation. 
The existence of a tight distribution in colour
when the universe was at one quarter of its current age implies that
mechanisms that have not yet been implemented in current
galaxy formation scenarios long ago began to shape the star formation history
of red cluster galaxies.
}


   \maketitle

\section{Introduction}

In $\Lambda$CDM cosmology there may be multiple epochs of galaxy formation
depending on the phenomenon one concentrates on. Indeed, we expect that
galaxies may be younger than their stars, because the mass is assembled hierarchically
via the progressive merger of lower luminosity objects. The two epochs of
star formation and mass assembly then may yield constraints on the mechanisms
of galaxy formation and evolution. Clusters of galaxies are particularly
important in this context, because they provide volume-limited samples of
galaxies observed at the same cosmic epoch and in an environment corresponding
to the highest density peaks in the dark matter distribution at each
redshift. It is therefore possible to simultaneously constrain the star
formation history (via galaxy colours) and mass assembly history (via the
luminosity function) and probe galaxy formation to high redshift, with 
relatively simple and inexpensive observational campaigns rather than
the extensive spectroscopic surveys needed for field galaxies.

Current data show that the colour-magnitude relation of bright galaxies in 
clusters to $z=1.4$ is very tight and that it evolves passively, thereby implying 
that their stellar populations must have formed rapidly and at high redshift 
(e.g. Stanford et al. 1998; Kodama et al. 1998, Papovich et al. 2010 and
references therein).
Recently, Andreon et al. (2009) detected what seems to be the most distant 
cluster of galaxies to date (JKCS\,041 at $z_{phot}=2.2$)
by looking at over densities in red galaxies. In the
Chandra follow-up, the cluster 
is detected as an extended X-ray source with $T\sim7.4$
keV (Andreon et al. 2009; Andreon et al. 2011), 
which confirms that there is  a hot intracluster medium, present in
formed clusters and lacking in protoclusters. Deep near-infrared 
observations show that the cluster has a prominent red
sequence (Andreon \& Huertas-Company 2011a,b), locating
the cluster at $z_{phot}=2.20\pm0.11$.

Throughout this
paper, we assume the following cosmological parameters:
$H_0=71$ km s$^{-1}$ Mpc$^{-1}$, $\Omega_m=0.27$, and
$\Omega_\Lambda=0.73$. Magnitudes are in the AB system. 

\section{Data} 


JKCS\,041 is in the area covered by  
CFHTLS deep survey 
and by  WIRDS follow-up in the infrared filters
($J, K_s$) 
(catalogues are available on the Terapix site, see also
Andreon \& Huertas-Company 2011a,b). 
More precisely, we used the catalogue generated using
$K_s$-band as detection image and the other bands ($z'$, $J$ and $H$)
in analysis mode. 
Images have been spatially resampled and filtered during the combining
phase, which is why the photometric errors computed by Source Extractor
(Bertin \& Arnouts 1996) are 
underestimated. We assumed an error correction factor of 1.5 for optical
images, in agreement with Ilbert et al. (2006) using the very
same optical images, and a factor 2.5 for the more finely 
resampled NIR images, in agreement with the analysis of
Raichoor \& Andreon (in preparation) of the very same images.

In this paper, we only consider objects brighter than $K_s=23.0$ mag.
The claimed 50~\% completeness for point sources is 1.9 mag deeper
(Bielby et al. 2010).
Figure 1 shows the colour-magnitude relation 
for galaxies within an ellipse
of parameters $(ra,dec,a,b,PA)=(36.687745, -4.6940845,$ $85.6", 55.8", 330)$,
chosen to maximise the contrast between cluster and background galaxies.
In this figure only, blue ($z'-J<1.75$ mag) galaxies in the SSW octant
has been flagged, because this blue part of the diagram is
contaminated by a nearby structure (see Andreon \& Huertas-Company 2011a for
details).
The cluster red sequence clearly stands out.
In order to measure the colour spread, and from it the age spread,
the colour range to select red galaxies should be large,
to avoid very similar colours for
the selected galaxies, and as a
consequence spuriously synchronized star formation histories, because of
the too stringent adopted colour cut (see Andreon et al. 2008 for a discussion).
We therefore define as red galaxies those with $z'-J>1.75$ mag
and these galaxies are plotted in Fig. 1 as solid points.
Bluer galaxies are manifestly too blue to be red sequence galaxies 
in $z'-J$ (they are $> 0.22$ away from the red sequence, i.e. $>2.9$ times
the red sequence thickness, see also Fig. 2 and sec 4).    
There are 23 galaxies satisfying the constraint above, of which 3.6
are expected to be background galaxies, based on the counts measured in
a region of 0.7 deg$^2$ all around the cluster. 
The $z'-J$ colour has been adopted because it
offers the best way of
selecting red {\it cluster} galaxies than any other pair of filters 
with larger wavelength baseline. 

\begin{figure}  
\centerline{\psfig{figure=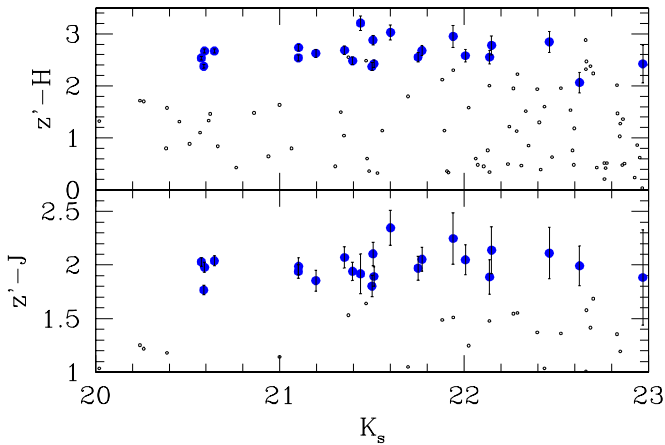,width=9truecm,clip=}}
\caption{colour-magnitude plot of galaxies in the JKCS\,041 direction.
Red sequence ($z'-J>1.75$ mag) galaxies are plotted with solid (blue) points.
Error bars are marked for these galaxies only, to avoid crowding. 
} 
\label{fig:cm} 
\end{figure}

Fig. 2 shows the colour distribution of the galaxies in the
JKCS\,041 direction (points with heuristic error bars), and
in the background direction, the latter computed from a 0.7 deg$^2$
all around the cluster and normalized to the considered cluster
solid angle (bottom solid line). In the left panel, all galaxies
in the JKCS\,041 line of sight are displayed,
whereas the $z'-J>1.75$ mag selection is applied in the next
two panels. These show an isolated peak in the colour distribution,
of about 0.18 mag width in $z'-H$ and $J-K_s$. 
The colour distribution of the background galaxies
(see bottom solid line) is, as expected, broader and negligible.

\begin{figure}  
\centerline{\psfig{figure=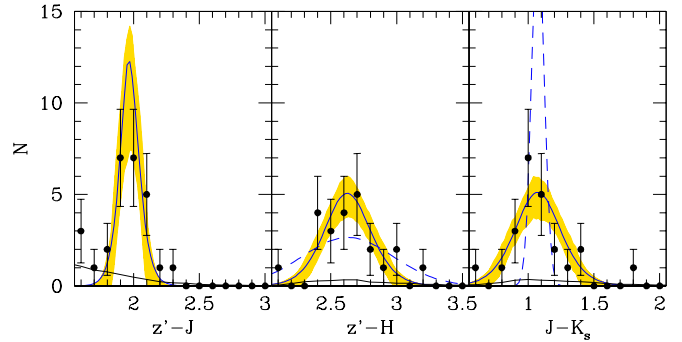,width=9truecm,clip=}}
\caption{colour distribution of galaxies in the JKCS\,041 direction (points)
and in a control region all around the cluster, normalized to the cluster
solid angle (lower, black, curve). The (blue) curve with shaded 
68 \% (highest posterior credible) intervals is
a Gaussian fit to the data. Approximated (i.e. $\sqrt{n}$) error bars
are shown for display
purpose only. The dashed blue Gaussian lines show the
colour spread expected for a 250 Myr (central panel)
or 50 Myr (right panel) age spread. 
} 
\label{fig:colourhisto} 
\end{figure}

\section{Analysis}

The spread across galaxies in star formation history can be
inferred from the width of the above colour distributions.
The expected spread for a population with an age spread of
50 (250) Myr is plotted in the right (central) panel (dashed blue lines), 
and the age spread for cluster galaxies should be somewhere in between.

To constrain the star formation history of  
red galaxies via their colour homogeneity, 
we were inspired by Bower, Lucey \& Ellis (1992) and following
work, e.g. by Andreon (2003), and Andreon et al. (2009). a) We assumed 
that the $i^{th}$ galaxy forms stars at the
time $t_{i}$, i.e. can be modelled by a single stellar population (SSP, 
but see later for other choices of star formation histories). b) These individual
formation times are spread around the mean star formation age $t$ with a spread $\sigma$
(i.e. $t_i \sim \mathcal{N}(t,\sigma^2)$).
c) With a stellar population synthesis model, we inferred 
the colour of each modelled stellar population. d) By a Markov Chain
Monte Carlo we updated the parameters $t$ and $\sigma$ until the model colour
distribution matched the observed colour distribution. In this step, we
account for the heteroscedastic errors, which broaden the 
colour distribution by an amount that depends on each galaxy. We assumed
a uniform prior for $t$, zeroed at $t-\sigma<0$ Gyr (i.e. galaxies form {\it 
before} observing them) 
and above $t-\sigma>2.54$ Gyr (about $z\sim8$, only about 0.5 Gyr
away from the Big Bang). We assume an uniform prior
for the spread $\sigma$, zeroed at $\sigma<0$
because the spread is a positively defined quantity.  In our
reference analysis, we used the 2007 version of Bruzual \& Charlot
(2003) SSP of solar metallicity and 
Chabrier initial mass function.

As mentioned several 
times in the literature, the colour evolution is highly sensitive to systematic
uncertainties in the stellar population synthesis model (Worthey 1994), and so it is preferable to
rely only on differential measurements, i.e. on the colour scatter. We therefore allow stellar population synthesis models
to have an unknown zero-point (colour) error as large as $\pm0.4$ mag by adding
a parameter to the model colour with uniform prior zeroed outside $\pm0.4$ mag.

Given the good quality of the data, the posterior is independent of the prior
(i.e., results are robust against the prior choice).

\section{Results} 

By using $J-K_s$ (rest-frame $U-R$) alone, we find a spread of
stellar ages of $190\pm40$ Myr. A similar
result, $130\pm50$ Myr, is found using $z'-H$ alone (rest-frame $UV-V$). 
Using both colours at once, 
we find a spread of $160\pm30$ Myr. The constraint
basically comes from the fact that higher/lower 
values imply too broad/narrow distributions
in colour, independently of the value of the average star formation age 
$t$. The (posterior) distribution probability of the spread $\sigma$ in 
star formation age is shown in Fig. 3.

We succeeded at measuring a small scatter in stellar ages because the colour
change per Gyr is much higher for younger populations (i.e.
at high redshift) than for older ones (i.e. at low redshift).
For example, at low redshift, a 0.02 mag colour scatter translates into a 400-700
Myr scatter, depending on the adopted plausible age of the stellar
population. By going to a high redshift (younger ages), colours change faster, by
0.1 mag per 100 Myr, therefore the same 0.02 mag colour scatter corresponds,
roughly, to a 20 Myr spread in the star formation age.
Furthermore, there is little room for choosing the stellar age, because it is
tight bounded by the Big Bang.

\begin{figure}
\centerline{\psfig{figure=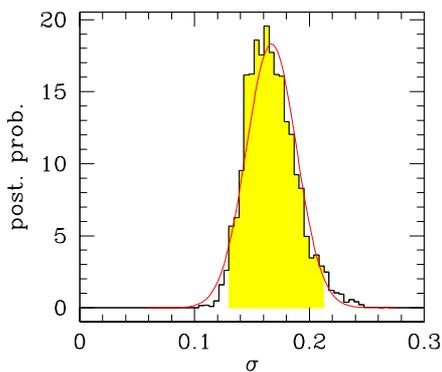,width=6truecm,clip=}}
\caption{Posterior probability distribution of the spread $\sigma$
of the stellar age of red JKCS\,041 galaxies in Gyr. The shading
highlights the 95 \% (highest posterior) interval and the red curve is the
Gaussian approximation.} 
\label{fig:boh}
\end{figure}

To measure the robustness of our result on assumptions, 
we removed one of them at the time. First, 
we replaced the SSP model with a model with an
exponential decreasing star formation, with $\tau=0.1$ Gyr. 
We found a spread of $150\pm20$ Myr, less than half a sigma different 
than adopting SSP's. We had little freedom in choosing much larger 
$\tau$ models, because those with $\tau>0.5$ 
have colours that are so blue that they
do not match the observed colours, even when allowing for
an (implausibly large) model systematic
error of 0.4 mag. Adopting a $\tau=0.3$ Gyr, we got a spread
of $230\pm50$ Myr, but with tension between the adopted model 
and the data, because the model is still not red enough to match
the data, even when allowing for the 0.4 mag error.

If we use SSP models from Maraston et al. (2005), which
use the fuel consumption theorem instead of the isochrones used
in Bruzual \& Charlot (2003), we find an identical result:
$160\pm40$ Myr. 
If we use a different combination of colour indexes   
($z'-J$, $J-H$, $J-K_s$), we find $170\pm25$ Myr, where the error only
approximatively accounts for the dependent nature of the three
considered colour indexes, again in full agreement with the reference
analysis.
If we adopt a value of 2.0 for the correction factor for errors in
all bands,  we find a spread of $180\pm30$ Myr, in full agreement
with the reference analysis.
If the cluster redshift were $z=2.0$, i.e. two sigmas below
the measured value of $z=2.20 \pm 0.11$, 
then the age spread would be  $160\pm25$ Myr, in full agreement
with the reference analysis.
If we adopt SSP models with twice higher metallicity, then the age-colour
relation steepens a bit, and, as a consequence, a narrower
spread in ages is needed to reproduce the same observed
colour spread. In fact, we find $120\pm 25$ Myr,
still in line with our
reference analysis, since it is only $1.0$ combined sigmas away from it.

Our analysis goes from parameters (e.g. age spread) to the data (measured 
colour spread) in a single step. It may, however, be
interesting to take the inverse approach, by
inferring the intrinsic scatter of the colour distribution and
from it the age scatter.
We ignore the background contribution (see  Andreon et al. 2008 if
one chooses not to ignore it), and we fit the colour and its 
dispersion accounting for errors with uniform prior on parameters
(zeroed for negative, unphysical, dispersions). We find
a colour dispersion (red sequence tickness) of $0.18\pm0.04$ mag for $z'-H$ and $J-K_s$, and
$0.08\pm0.02$ mag in $z'-J$. The fits are depicted in Fig.~2.
This colour scatter may be inverted, as in Bower, Lucey \& Ellis (1992)
and many works since them, into a $\sim 180$ Myr age spread. This 
agrees with our derivation, although we prefer
the forward (adopted) approach, instead of inverting the data.

If anything,
the measured value is a slightly overestimation of the true age spread,
because we have neglected the possibility of a scatter in metallicity or
dust content, secondary possibly stochastic minor episodes of star
formation superposed on an essentially single stellar population, as well
as the (minor) effect of background galaxies among red galaxies and
the negligible effect of the colour--magnitude slope. 
All these terms tend to increase the expected observational scatter
at a given age scatter $\sigma$, unless there is
a tuned covariance between them to keep the observed colour scatter
small.

\section{Conclusions} 

We found an age spread of $160\pm30$ Myr, independently of the assumptions
used to infer ages from a colour scatter, namely the
absolute colour of the stellar populations, the 
synthesis population code used, the way the 
star formation history of each galaxy is modelled, the adopted
metallicity, the choice of the colour index, and the precise
redshift of the cluster and the approach (forward or inverse) used
in analysing the data.
We succeeded at measuring such a small scatter in stellar ages because the colour
change per Gyr is much higher for younger populations (i.e.
at high redshift) than for older ones (i.e. at low redshift), and because
the range of possible ages is much reduced at high redshift,
as we are approaching the Big Bang.

The extreme synchronicity of red galaxies
in JKCS\,041 is in marked disagreement with the idea that galaxies
in clusters form by a hierarchical assembly of smaller units, unless
this occurs a) over such a short time that it is essentially equivalent
to an (un-physical) monolithic assembly, or b) by gas-free merging. 
In fact, a hierarchical
assembly has some stochastic variations in the merging histories, which
induce differences in colours because some merging events involve
gas. It is worth noticing that
the Menci et al. (2008) rendition of the hierarchical paradigm predicts
the disappearance of the red sequence in clusters
at redshift around 1.5 to 2.0.

In the $z=2.2$ JKCS\,041 cluster, the red sequence is already populated
by a homogenous population of galaxies with extremely synchronised 
stellar ages, which implies that mechanisms not yet implemented in current
galaxy formation scenarios long ago began to shape the star formation histories
of red cluster galaxies.

\begin{acknowledgements} 
I thank Anand Raichoor for useful discussions  and Claudia Maraston for
her models in electronic form. This work is based on observations with
MegaPrime/MegaCam\footnote{The full text acknowledgement is at
http://www.cfht.hawaii.edu/Science/CFHLS/cfhtlspublitext.html} and 
WIRCAM\footnote{The full text acknowledgement is  at
http://ftp.cfht.hawaii.edu/Instruments/Imaging/WIRCam/WIRCamAcknowledgment.html}
at CFHT. 
\end{acknowledgements}

{}


\begin{thebibliography}{}

\bibitem[{Andreon}{2001}]{2001ApJ...547..623A} 
Andreon S., 2001, ApJ, 547, 623 

\bibitem[{Andreon}{2003}]{2003A&A...409...37A} 
Andreon S., 2003, A\&A, 409, 37 

\bibitem[Andreon(2006)]{2006A&A...448..447A} 
Andreon, S.\ 2006a, A\&A, 448, 447 

\bibitem[Andreon(2006)]{2006MNRAS.369..969A} 
Andreon, S.\ 2006b, MNRAS, 369, 969 

\bibitem[Andreon(2008)]{2008MNRAS.386.1045A} 
Andreon, S.\ 2008, MNRAS, 386, 1045 

\bibitem[{Andreon \& Huertas-Company}{2011}]{2011A&A...526A..11A} 
Andreon S., Huertas-Company M., 2011a, A\&A, 526, A11 

\bibitem[{Andreon \& Huertas-Company}{2011}]{} 
Andreon S., Huertas-Company M., 2011b, Mem SAIt, in press (arXiv:1012.3051A)

\bibitem[Andreon et al.(2006)]{2006MNRAS.372...60A} 
Andreon, S., Cuillandre, J.-C., Puddu, E., \& Mellier, Y.\ 2006, MNRAS, 372, 60 

\bibitem[Andreon et al.(2009)]{2009A&A...507..147A} 
Andreon, S., Maughan, B., Trinchieri, G., \& Kurk, J.\ 2009, A\&A, 507, 147 

\bibitem[{Andreon, Trinchieri, \& Pizzolato}{2011}]{2011MNRAS.tmp...73A} 
Andreon S., Trinchieri G., Pizzolato F., 2011, MNRAS, in press (arXiv:1012.3034) 

\bibitem[Andreon et al.(2008)]{2008MNRAS.385..979A} 
Andreon, S., Puddu, E., de Propris, R., \& Cuillandre, J.-C.\ 2008, MNRAS, 385, 979 

\bibitem[{Bertin \& Arnouts}{1996}]{1996A&AS..117..393B} 
Bertin E., Arnouts S., 1996, A\&AS, 117, 393 

\bibitem[{Bielby et al.}{2010}]{2010A&A...523A..66B} 
Bielby R.~M., et al., 2010, A\&A, 523, A66 

\bibitem[{Bower, Lucey, \& Ellis}{1992}]{1992MNRAS.254..601B} 
Bower R.~G., Lucey J.~R., Ellis R.~S., 1992, MNRAS, 254, 601 

\bibitem[Bruzual \& Charlot(2003)]{2003MNRAS.344.1000B} 
Bruzual, G., \& Charlot, S.\ 2003, MNRAS, 344, 1000 

\bibitem[{Ilbert et al.}{2006}]{2006A&A...457..841I} 
Ilbert O., et al., 2006, A\&A, 457, 841 

\bibitem[{Kodama et al.}{1998}]{1998A&A...334...99K} 
Kodama T., Arimoto N., Barger A.~J., Arag'on-Salamanca A., 1998, A\&A, 334, 99 

\bibitem[{Maraston}{2005}]{2005MNRAS.362..799M} 
Maraston C., 2005, MNRAS, 362, 799 

\bibitem[Menci et al.(2008)]{2008ApJ...685..863M} 
Menci, N., Rosati, P., Gobat, R., Strazzullo, V., Rettura, A., Mei, S., 
        \& Demarco, R.\ 2008, ApJ, 685, 863 

\bibitem[Papovich et al.(2010)]{2010ApJ...716.1503P} 
Papovich, C., et al.\ 2010, ApJ, 716, 1503 

\bibitem[{Stanford, Eisenhardt, \& Dickinson}{1998}]{1998ApJ...492..461S} 
Stanford S.~A., Eisenhardt P.~R., Dickinson M., 1998, ApJ, 492, 461 

\bibitem[{Worthey}{1994}]{1994ApJS...95..107W} 
Worthey G., 1994, ApJS, 95, 107 

\end{thebibliography}
\end{document}